\documentclass[11pt]{article}
\usepackage{moriond,epsfig}

\def\be{\begin{equation}}
\def\ee{\end{equation}}
\def\bea{\begin{eqnarray}}
\def\eea{\end{eqnarray}}

\begin{document}
\vspace*{4cm}
\title{INSTRUCTIONS FOR PRODUCING A CAMERA-READY MANUSCRIPT}

\title{STRINGS, BRANES AND COSMOLOGY:\\ WHAT CAN WE HOPE TO LEARN?}

\author{ C.P. BURGESS }

\address{Department of Physics \& Astronomy, McMaster University\\
1280 Main Street West, Hamilton, Ontario, Canada L8S 4M1\\
{\it and}\\
Perimeter Institute for Theoretical Physics\\
31 Caroline Street North, Waterloo, Ontario, Canada N2L 2Y5.}

\maketitle\abstracts{This article briefly summarizes the
motivations for --- and recent progress in --- searching for
cosmological configurations within string theory, with a focus on
how much we might reasonably hope to learn about fundamental
physics from precision cosmological measurements.}

\section{Why String Cosmology?}

The last few years have seen a number of scientific gatherings
which have brought together string theorists and cosmologists in a
way which would have been unheard-of only a few years ago,
stimulated by a relatively recent convergence of interest between
these two fields. This convergence is interesting in its own
right, promising as it does to relate the laws of nature at the
smallest of distances to the behaviour of the universe as a whole,
as seen writ large across the sky by contemporary cosmologists.
But the possibility that there should be such a connection also
contains within itself a puzzle, to do with why it is possible for
these two fields usefully to inform one another at all.

\subsection{Why doesn't string theory decouple from cosmology?}

What is so puzzling about a connection between string theory and
cosmology? The puzzle inherent in this field is intimately tied up
with its promise: the potential it holds out for a connection
between the workings of nature on the smallest of distance scales
and the properties of the universe at large. Such a connection is
intrinsically puzzling because of a fundamental property of
nature, which might be called the {\it Principle of Decoupling}.

\begin{quote}
{\it The Principle of Decoupling:} Although the world comes to us
in many scales, these scales can each be understood on their own
terms since their properties do not depend strongly on all of the
details of the physics of other scales.
\end{quote}

For example, we know that atoms are built from constituents
(electrons and nuclei) which are much smaller than the atoms
themselves, and some of these constituents (nuclei) themselves
consist of still smaller things like quarks and gluons. But a
detailed understanding of atomic properties ({\it i.e.} the
spectra and chemistry of atoms) depends only on gross properties
of their constituents (like the nuclear mass and charge). In
particular, it does {\it not} depend on any of the complicated
details of how they are constructed from their own constituents.
Historically, this is why it was possible to understand atomic
physics before having a complete understanding of nuclear physics.
Indeed, this property of nature could be argued to be an important
part of the reason why Science is possible at all, since it shows
why we can hope to understand part of what is going on in nature
without having to understand everything at once.

This gives rise to the puzzle: given the decoupling of scales in
nature, how can cosmology --- the understanding of the properties
and behaviour of the largest objects known --- possibly depend on
the details of string theory --- our best candidate for a theory
of nature at the very smallest of scales? After all, we don't have
meetings at which condensed-matter physicists or atomic physicists
expect to learn much that is useful from string theorists. These
meetings don't take place because string theory is likely to get
right all of the details of atomic and condensed matter physics
provided only that it predicts the existence of electrons and
nuclei, and it gets right the laws of electromagnetism (as
expressed by QED). For this reason there is little to be gained by
comparing string predictions with detailed measurements of
condensed matter phenomena.

There seem to be three reasons why string theory can usefully
inform cosmology, and vice versa.

\begin{enumerate}
\item{\it Access by cosmology to very high energies:} This is the
traditional reason for the decades-old development of a fruitful
interface between astrophysics and particle physics. Some
astronomical systems (like active galactic nuclei or
ultra-high-energy cosmic rays) can involve physical processes
involving astronomically large energies, whose understanding
requires knowing how high-energy elementary particles behave.

The same can be true for cosmology because we know that the
observable early universe is well described by the Hot Big Bang
model, but only if special initial conditions (homogeneity,
isotropy, flatness, and a spectrum of primordial density
fluctuations) are chosen before the earliest epoch
(nucleosynthesis) for which we have direct observational evidence.
Although nuclear physics seems to suffice for understanding
nucleosynthesis, particle physics is required in order to
understand the origin of the special initial conditions. In
particular, the extremely high energies associated with string
theory are very likely to be important if these initial conditions
are explained by a very early epoch of inflationary universal
expansion.\cite{Inflation}
\item {\it Dependence on UV-sensitive properties:} Cosmology is
unusual because the vast majority of cosmological models rely for
their phenomenological success on properties which are notoriously
sensitive to microscopic details. For example specific models of
Dark Energy \cite{DEReview} or inflation
\cite{InflationReviews,InflationBooks} often depend on the
existence of very shallow scalar potentials which give rise to
extremely light scalar masses, $M_\phi \le H$, where $H$ is the
Hubble scale at the epoch of interest. But scalar masses are
famously difficult to keep from getting large contributions when
the short-distance (UV) sector of the theory is integrated out. To
take an extreme case, most Dark Energy models require $M_\phi <
10^{-33}$ eV, while it is difficult to make the contribution to
$M_\phi$ due integrating out a particle of mass $m$ smaller than
$\delta M_\phi \sim m^2/M_p$, where $M_p = (8 \pi G)^{-1/2} \sim
10^{18}$ GeV. This correction is already larger than $M_\phi$ for
$m > 10^{-3}$ eV, and so is many orders of magnitude too large
even for the electron, for which $m_e \sim 5 \times 10^5$ eV.
\item{\it Difficulty of modifying gravity on long distance
scales:} Much of the evidence for the existence of exotic matter
(like the scalar particles just mentioned) in cosmology is based
on inferences which assume General Relativity is the correct
theory of gravity. But General Relativity has never been
experimentally tested over distances as large as required for
cosmological applications, and this observation has led many
people to try to avoid the need for exotic matter by instead
appropriately modifying gravity at long distances. Some
phenomenological success can be achieved along these lines,
provided one is judicious in the modifications which are made.

However, what this line of argument misses is that it is {\it
extremely} difficult to embed any modification of gravity at long
distances into any kind of a sensible theory of short-distance
physics. This is because we now know that General Relativity is
the most general kind of interaction which a massless spin-two
particle can have which is consistent with very general principles
(like special relativity, stability and unitarity), and as a
result we have a very general understanding as to {\it why}
General Relativity provides a good description of
gravity.\cite{WhyGRa,WhyGRb} So far as is known, it seems very
likely that {\it any} sensible theory of gravity must look in the
far infrared (IR) like a combination of scalar fields and gauge
fields interacting with General Relativity, and there is no
compelling theory which is both consistent with measurements in
the solar system and in astrophysics and yet also observably
different from gravity at very long distances in a
phenomenologically successful way.\cite{DGP} This indicates that
consistency issues at short distances provide an important clue as
to what is possible to entertain as a description of nature over
long distances.\footnote{It must be emphasized that because these
issues deal with {\it long-distance} problems, they may be
unambiguously addressed using current knowledge --- using standard
Effective Field Theory techniques \cite{GREFT} --- and in
particular need not await an eventual `final' theory of Quantum
Gravity, as is sometimes argued.}
\end{enumerate}

For the above reasons there is at present an unusual opportunity
at the interface between cosmology and microphysics, which
provides a real chance for learning something important about
nature. The opportunity arises because the very success of
cosmological models relies in detail on properties (like shallow
scalar potentials) which we know to be extremely sensitive to the
details of short-distance physics. Furthermore, it is not generic
that these microscopic details provide phenomenologically
successful models for cosmology. The condition that a model {\it
both} provide successful phenomenology {\it and} be sensibly
embedded into microscopic physics is very strong, making the
finding of examples which do both a worthwhile exercise.
Furthermore, as we now argue, there is an opportunity for
information to flow in {\it both} directions, with potential
theoretical insights for both string theory and cosmology.

\subsection{What is useful for cosmologists}

We first ask how short-distance physics can be useful for
practical cosmologists interested in understanding observational
data. The utility here comes from the observation that
cosmological observations (marvellously precise though they are)
are likely to remain inadequate into the foreseeable future for
unambiguously differentiating amongst the many competing
phenomenological cosmological models.\cite{CosmoInadequate}

However cosmological observations provide only part of the clues
as to what is going one. We must also weed out those models which
do not make sense when embedded into more microscopic theories,
and it is the interplay between these two kinds of constraints
which makes the exercise theoretically constrained. In practice
this means ruthlessly rejecting those models of cosmology which
predict low-energy ghosts, instabilities or violations of the
experimental tests of gravity within the solar system or for
binary pulsars. Such a restriction dramatically reduces the number
of models which require more detailed scrutiny.

\subsection{What is useful for string theorists}

The information exchange between string theory and cosmology is
likely also to be of use to string theorists, for the following
reason. String theory involves an enormous number of degrees of
freedom and so may be expected to enjoy an equally enormous number
of solutions. A precise counting of how many solutions there might
be requires an understanding the form of the potential which
stabilizes the many fields of the theory, but recent progress
\cite{GKP,sethi1,KKLT} in computing this potential for some types
of string vacua indicates there to be more than $10^{100}$ such
vacua. A central question for string theorists is to find which
solutions can describe the universe around us, and to understand
why the universe should end up being described by these solutions
rather than by the many other possible solutions.

Cosmology may help this process in two ways. First, cosmology can
help to {\it find} string vacua with acceptable phenomenology. It
can do so because the direct examination of various vacua is
impractical, given the likely enormous number of solutions which
exist. What can be useful when looking for potentially realistic
vacua, however, is the identification of low-energy {\it modules}
which capture one or another of the phenomenologically desirable
features required to describe our low-energy world. For instance,
these could include modules for ensuring an acceptable particle
spectrum; a mechanism for understanding the electroweak hierarchy,
and so on. Some of these modules can involve cosmology, such as by
demanding the existence of an early inflationary phase; a
candidate for dark matter; or an understanding of the observed
features of the dark energy density.

The second useful role cosmology might play for string theory is
by providing potentially measurable signals for comparison with
experiments. Recall that the existence of an enormous number of
vacua makes the extraction of a theory's predictions much more
complicated, since the properties of each vacuum provide in
principle a separate set of predictions for what might be found
around us. The most likely way in which such a theory will be
tested in practice is through its statements about the {\it
correlations} of the properties to be found about any particular
vacuum, along the lines of: ``Any vacuum which has property X must
also have property Y''. For instance, X might be the statement
``contains the standard model gauge group, and Y might be ``has 3
generations of chiral fermions''.

Cosmology can usefully contribute to the kinds of statements, X
and Y, since it is plausible that our understanding of why the
universe is the way it is now will depend on our understanding of
where it has been in its past. For instance X or Y might include
``has at least 60 $e$-foldings of inflation'', or ``has
such-and-such a relic abundance of cosmic strings''. In
particular, one can hope that the class of string solutions which
give a reasonable description of cosmology might also lead to a
restricted class of particle physics properties to be compared
with laboratory experiments.

\section{Branes and naturalness}

An important way in which string theory has influenced thinking
about more phenomenological issues can be traced to the discovery
of branes.\cite{Dbranes} This discovery has radically changed the
kinds of low-energy implications which the vacua of the theory can
have, and this has in turn led to a number of important new
insights into the nature of the various `naturalness' problems
which arise within the effective theories relevant for
phenomenology.

\subsection{Why are branes important?}

The main reason why branes have provided new insights into
low-energy naturalness problems is because the study of the
low-energy properties of vacua containing branes has identified a
number of important (and overly restrictive) hidden assumptions
which had been hitherto made regarding what is possible for the
low-energy limit of a sensible high-energy theory.

The identification of such assumptions is crucial for naturalness
problems, because these problems in essence amount to statements
like: ``a broad class of low-energy theories (obtained by
integrating out heavy modes in some fundamental theory) have a
generic property X, which is not observed to be true in nature.''
Property X here might be: ``has a Higgs mass similar in size to
the Planck mass'', or ``has a large cosmological constant.'' It is
crucial to know when these unwanted but generic properties depend
on hidden assumptions, since these may prove to be unwarranted and
so may be the loopholes through which nature evades the problem.

For instance, a very important hidden assumption which the study
of branes has identified is the assumption that all interactions
`see' the same number of spacetime dimensions. This assumption is
violated, for instance, if particles like photons arise from open
strings, which at low energies are localized on the branes on
which such strings must end. In this case photons must propagate
only within the dimensions spanned by the branes, while gravitons
can move throughout the full extra-dimensional environment. Among
the suggestive new insights which have emerged in this way are:
\begin{enumerate}
\item {\it A Lower String Scale:} The string scale need not be
close to the Planck scale,\cite{StringScaleOpen} opening up
interesting new possibilities for understanding the electroweak
hierarchy with the string scale being associated with the
intermediate scale \cite{IntScale} or the TeV scale.\cite{ADD}
\item {\it Large Extra Dimensions:} A possibility which is related
to (but not identical with) having the string scale at the TeV
scale is that extra dimensions can be much larger than had been
thought, being potentially as large as micron size.\cite{ADD}
\item {\it Decoupling 4D Vacuum Energy from 4D Curvature:} In four
dimensions a large vacuum energy is identical with a large
cosmological constant, and so also with a large 4D curvature.
(This connection underlies the cosmological constant problem since
the curvature is observed to be small while the vacuum energy is
expected to be large.) Higher-dimensional brane solutions show
that this connection need not survive to higher dimensions, where
large 4D energies can co-exist with flat 4D
geometries.\cite{5DSelftune,6DSelftune}
\item {\it Non-locality:} Locality is normally automatically
ensured for effective theories because these theories are defined
by integrating out only very heavy states. However since brane
constructions can allow extra dimensions to be large compared with
particle-physics length scales, the effective theories which
result can admit a restricted form of nonlocality. They can do so
because the observable particles might now be identified as those
living on a collection of branes, rather than simply in terms of a
low-energy limit. For instance, interactions which are obtained by
integrating out modes which are {\it not} heavy compared with TeV
scales --- such as bulk Kaluza Klein (KK) states --- can mediate
nominally nonlocal correlations into the remaining fields.
\end{enumerate}

These possibilities show why string theory may have the potential
to teach us a considerable amount about how to think about any new
physics which might be encountered in future observations, even if
the string scale should turn out to be much higher than the
energies being directly probed in these experiments. The fact that
string theory makes it plausible that the particles we observe
might be trapped on branes within extra dimensions, and that this
possibility changes how we think about general naturalness issues,
makes it worthwhile to take the possibility of brane localization
very seriously.

\section{String Inflation}

Inflation is the simplest application of string theory to
cosmology to motivate, because it could easily involve energy
scales which are so high that they could plausibly directly probe
string-related physics. Furthermore, recent precision measurements
\cite{WMAPInflation} of the properties of the cosmic microwave
background radiation (CMBR) have accumulated impressive evidence
supporting the existence of an early inflationary epoch. One of
the pleasures of this particular meeting at Moriond was the very
recent announcement of the most precise such measurements
\cite{WMAP3} to date.

Observations of the CMBR are only sensitive to essentially three
numbers in any slow-roll inflationary
model:\cite{InflationReviews} the inflationary Hubble scale,
$H_{\rm inf}$, and its first and second logarithmic derivatives
with respect to the scale factor, evaluated at `horizon exit'
({\it i.e.} the moment when observable scales cross the Hubble
scale). (It is conventional to describe these latter two
derivatives in terms of two small dimensionless slow-roll
parameters, $\epsilon$ and $\eta$.)

In principle these three parameters provide one relationship
amongst the four observables defined by the amplitude and spectral
tilts of the primordial spectrum of scalar and tensor
perturbations to the metric, although the full power of this
prediction is difficult to fully exploit until tensor fluctuations
are detected. In the meantime, one may instead constrain
$\epsilon$ and $\eta$ from measurements of the scalar spectral
tilt, $n_s$, as measured from the fluctuations in the CMBR, and
from upper limits on $r$, defined as the ratio of the amplitude of
tensor fluctuations to the amplitude of scalar fluctuations.

At present, current measurements are only now starting to be able
to distinguish between the predictions of broad classes of models.
Three classes of models which may be distinguished in this way
are:\cite{WMAPInflation,WMAP3}
\begin{enumerate}
\item {\it Large-Field Models,} for which $\epsilon$ and $\eta$
vary inversely with the value of the inflaton field: $\propto
(M_p/\varphi)^p$, for some $p > 0$;
\item {\it Small-Field Models,} for which $\epsilon$ and $\eta$
are proportional to a positive power of the value of the inflaton
field: $\propto (\varphi/M_p)^p$, for some $p > 0$;
\item {\it Hybrid Models,} for which field evolution at the end of
inflation involves at least a two-dimensional field space, and for
which the slow-roll parameters depend on parameters in the
potential which govern the couplings between these fields.
\end{enumerate}

Varying the various parameters in these models leads to
predictions which fill regions of the observable $r-n_s$ plane. In
the limit where $H_{\rm inf}$ is essentially constant during
horizon exit ({\it i.e.} $\epsilon, \eta \approx 0$), all
slow-roll models approach the scale-invariant point, corresponding
to an unobservably small amplitude for tensor modes and a
precisely Harrison-Zeldovich (HZ) spectrum: $(r,n_s) = (0,1)$. But
each of the above classes tends to sweep out a different region of
predictions within the $r-n_s$ plane, all of which overlap near
the scale-invariant HZ point. In particular, the bulk of
small-field models (although not all) tend to prefer $n_s < 1$,
while the bulk of hybrid models (although not all) prefer $n_s >
1$. What is exciting about the latest CMBR observations
\cite{WMAP3} is that they are now beginning to exclude the HZ
point which is common to all classes of models, and as a result
are beginning to provide observationally-justifiable preferences
amongst these models.

\subsection{Why embed inflation within string theory?}

Given the few quantities to which observations are sensitive, the
skeptical reader might reasonably wonder whether it is premature
to invest considerable effort in finding inflationary evolution
within string theory. There are nevertheless several good reasons
for doing so. In particular, inflationary models must be embedded
into a fundamental theory like string theory in order to
understand the following issues:
\begin{enumerate}
\item {\it Naturalness:} Are the choices made in order to obtain
acceptable values for $H_{\rm inf}$, $\epsilon$ and $\eta$
inordinately sensitive to short-distance (UV) effects, or must
they be finely-tuned in order to achieve sufficient inflation?
(And if anthropic arguments are used to explain these
tunings,\cite{landscape} what assigns the probabilities
\cite{VacuumStatistics} which must be used in order to have an
adequate explanation?)
\item {\it Reheating:} At the end of inflation how does the energy
associated with inflation get converted into observable heat (as
is required in order to launch the present-day Hot Big Bang
epoch)? As anyone who lives in a cold climate knows: a warm house
requires {\it both} an efficient furnace and good insulation.
Likewise, for inflation it is not sufficient for there to be a
channel for coupling energy between the inflationary and observed
sectors, one must also show that too much energy is not lost into
any unobserved degrees of freedom. But this question cannot be
addressed without a proper theory (like string theory) of what are
{\it all} of the relevant degrees of freedom at inflationary
energies.\cite{StringReheat}
\item {\it Initial Conditions:} What justifies the choices which
are made for initial conditions {\it before} inflation? This
question can arise because part of the motivation for inflation is
to explain the unusual initial conditions of Hot Big Bang
cosmology. And inflation can itself require special initial
conditions for some kinds of inflationary models (such as for
hybrid models, for example). For such models the full microscopic
theory is required in order to understand the origin of these
initial conditions. Whether initial conditions really are a
problem depends on the type of model of interest, since for some
cases (like some large-field models) inflation can be an {\it
attractor} solution, inasmuch as it is the endpoint for a broad
class of initial conditions. Alternatively, for some cases (like
for small field models) one can instead appeal to eternal
inflation to explain why the inflating initial conditions might
come to dominate the later universe.\cite{LLM,SLV,topinf}
\end{enumerate}

\subsection{Why is string inflation so hard to find?}

Twenty years of experience has shown that it is quite difficult to
embed inflation into string theory in a controllable way. This is
somewhat paradoxical given that supersymmetric string vacua
provide so many massless scalar fields for which the corresponding
scalar potential is completely flat (and so for which $\epsilon =
\eta = 0$). The problem arises because a convincing case for a
slow roll requires a complete understanding of the potential for
these fields even after supersymmetry breaks. In particular one
must check that this stabilizing potential does not introduce new,
steep, directions into the potential along which the fields will
prefer to roll. Although a number of mechanisms were proposed over
the years taking advantage of supersymmetric flat
potentials,\cite{InflationReviews} the difficulty in reliably
computing supersymmetry-breaking effects, together with
cosmological problems with the resulting potentials in the few
calculable cases,\cite{overshoot} proved to be an obstacle to
further progress.

The introduction of branes proved to be the way forward for string
inflation, although the initial brane-brane proposal
 \cite{DvaliTye} also relied on supersymmetry for the flatness of its
potential (and so suffered from the same calculational
difficulties to do with supersymmetry breaking as did earlier
ideas). The decisive advantage of branes became apparent only much
later, for two reasons. First, it was realized that supersymmetry
breaking can become calculable, based on the mutual attraction of
a brane-antibrane pair \cite{BBbar,BBbar1} or branes at angles,
\cite{BBangles} leading to the brane-antibrane mechanism of
inflation. With calculability came an explosion of scenarios,
including models using $D3$ branes attracted towards $D7$
branes,\cite{d3d7} branes undergoing relativistic
motion,\cite{dbi} intrinsically stringy modes \cite{shamit},
extensions to M-theory vacua \cite{MTheory}, assisted inflation
using string axions \cite{Nflation} and more
--- see ref.~3 for more extensive references
than are possible here.

The second reason branes proved to be crucial for progess was the
insight they provided~\cite{GKP,sethi1} into the stabilization of
the many scalar fields of string vacua. Once the simplest vacuum
with all moduli stabilized was obtained,\cite{KKLT} its
combination with the brane-antibrane inflationary mechanism led to
the first inflationary scenario with a plausibly detailed string
pedigree.\cite{kklmmt,BCSQ} Shortly thereafter, variations on this
theme also led to the discovery of inflationary scenarios for
which it is the modulus describing the size of the extra
dimensions (and its axionic superpartner) which is the
inflaton.\cite{racetrack} Improved understanding of the potentials
which stabilize the moduli of string vacua, has allowed better and
better control over the approximations which are required in order
to establish inflation, enabling more detailed connections to be
made to the properties of explicit string
vacua.\cite{Conlon,BetterRacetrack}

\subsection{How natural is inflation in string theory?}

Now that some plausibly stringy inflationary models exist, how
fine-tuned do they appear to be? Although it is still a bit early
to draw definitive conclusions, since comparatively few corners of
field space have been explored to this point, some tentative
conclusions can be drawn. For instance, so far all of the
proposals but one (including in particular all of the
brane-antibrane scenarios) seem to require the same amount of
fine-tuning as do their field theoretical counterparts: {\it i.e.}
slow roll inflation requires parameters must be adjusted to within
a part in 100 or 1000. In the one example for which inflation
seems natural \cite{Conlon} it is a modulus, $X/M_p \sim
\ln(L/\ell_s)$, of the extra dimensions which is the inflaton.
(Here $L$ is the length of a cycle in the extra dimensions, $M_p$
is the 4D Planck mass and $\ell_s$ is the string length scale.) It
is natural because it takes advantage of a mechanism earlier
identified \cite{BMQRZ} in the field-theoretic limit, wherein the
inflaton potential takes the schematic form
\be
    V(X) = V_0  - A \, X^c \exp[- a \, (X/M_p)^c] + \cdots \,,
\ee
where $V_0$, $A$, $a$ and $c$ are constants, and the ellipses
represent terms which become important only as inflation ends.
Such a potential has a slow roll provided that $X \gg M_p$, but
the point is that this is {\it generic} to the domain of validity
of the effective theory in which this potential appears. It is
generic because $X \gg M_p$ corresponds to the condition $L \gg
\ell_s$, which is a prerequisite for describing the dynamics of
$L$ in an effective field theory. It is clearly of considerable
interest to see whether this example is representative, and if so
to identify reliably the regions of solution space where inflation
occurs so naturally.

\subsection{What kind of stringy effects can we hope to measure?}

Given that inflation appears to be possible in string theory, and
given the wealth and precision of current observations, can we
expect there to be any stringy `smoking guns' awaiting us in the
sky? Of course, a complete answer to this question must await a
proper exploration of the reheating problem in models containing
both inflation and a realistic standard model sector, since it is
only then that we can see how many stringy remnants might survive
into the late-time universe which we can observe. But three kinds
of broad conclusions about observable signals can already be
drawn.

\begin{enumerate}
\item {\it Remnant Cosmic Strings:} Within the brane-antibrane
inflationary mechanism inflation ends when the brane and antibrane
annihilate, and although not completely understood, it was
recognized from the beginning \cite{BBbar} that this process is
likely to generate an extremely rich spectrum of post-inflationary
remnants.\footnote{For high-dimension branes the cascade of
annihilations of these remnants might in some circumstances
provide a dynamical explanation for why 3-branes might be more
abundant at late times,\cite{BBbar} an idea which when
investigated in a fully cosmological context also predicts the
same for 7-branes.\cite{RK}} The key point for observational
purposes is that cosmic strings are special amongst these remnants
inasmuch as they can plausibly be produced with observable string
tensions and residual abundances,\cite{CosmicStrings1} although
whether they can live long enough to survive to the present epoch
is a somewhat model-dependent issue.\cite{CosmicStrings2}
\item {\it Observational Constraints Among Slow-Roll Parameters:}
In all of the calculations to date the conclusion that the
observed 4 dimensions inflate in a particular string (or
string-motivated) model is drawn using a low-energy 4D effective
field theory. As such, their direct predictions for the CMBR fall
within the category of predictions for 4D slow-roll inflationary
models. In particular, brane-antibrane models tend to fall into
the category of hybrid inflation models, with the earliest models
predicting \cite{BCSQ} a `blue' spectral index $n_s > 1$.
(Subsequent more detailed studies have shown this conclusion not
to be robust against adjustment of the details of the model, with
$n_s < 1$ being possible for some choices of parameters.\cite{cs})
By contrast, moduli-driven models, like those of the `racetrack'
type,\cite{racetrack,BetterRacetrack} are of the small-field type
for which $n_s < 1$ is more robustly preferred. (Indeed the most
recent of these \cite{BetterRacetrack} obtained $n_s = 0.95$ in
what was probably the last theoretical calculation not to be
biased by the most recent observations \cite{WMAP3} which favour
this value.) It is remarkable that the preference of the current
data \cite{WMAP3} for $n_s < 1$ already differentiates amongst
some of these models at a statistically significant level, by
differentiating amongst the classes of low-energy inflationary
field theories to which they give rise.\cite{HLS}

One might hope that string theory might be more predictive than
are the low-energy field theories which describe their effects at
low energies. For instance, this would occur if it happened that
not all of the three-dimensional inflationary parameter space ---
{\it i.e.} $H_{\rm inf}$, $\epsilon$ and $\eta$ --- of the 4D
field theories were generated by varying the underlying parameters
of the string models through all of their allowed values. This
would be an attractive possibility if it were true, since it might
permit a definitive test of string-based inflation by
observations. Unfortunately there is as yet no evidence that
string models do not explore the entire parameter space of 4D
inflationary slow rolls, although admittedly the parameter space
of the string-based models has not yet been extensively explored.
\item {\it Non-Decoupling Effects:} Everything known about string
theory is consistent with the dynamics around string vacua being
described at low energies by an appropriate effective field theory
--- although the occasional worry does get raised.\cite{Banks}
This allows a fairly robust analysis of the influence of
high-energy states on inflationary predictions for the CMB since
it is possible to analyze its effects in the effective field
theory limit. And this theory can be taken to be four dimensional
provided that the physics of interest around horizon exit is
itself four-dimensional. Since the cosmological backgrounds of
interest are time-dependent, care must be taken when performing
this analysis to keep track the additional conditions which arise
in this case for the validity of the effective-field theory
description.\cite{GREFT}

The results of such an analysis are interesting. First, one finds
that by far the majority of effective interactions do not perturb
the standard slow-roll inflationary predictions for the CMBR, with
a vast number of effects first arising at order $(H_{\rm
inf}/M)^2$, where $M$ is the relevant string (or KK) scale
describing the relevant high-energy physics.\cite{Shenker} This is
good news, since it ensures the robustness of the standard
predictions to high-energy string details. But there can be
exceptions to this statement, of two types.\cite{BCLH} One type
involves non-adiabatic time-dependent effects during the
$e$-foldings just before horizon exit. These effects can cause
deviations from the predictions of slow-roll inflation because
they violate the assumptions on which the slow-roll calculations
rely. Their existence is interesting since it motivates a careful
search within the observations for deviations from standard
slow-roll predictions.\footnote{It should be remarked
parenthetically that the expected deviations \cite{BCLH} can be
physically distinguished from those predicted by the more
speculative `transplanckian' effects \cite{TPI} which have been
much discussed recently.}

Alternatively, there can also be static effects \cite{BCLH} which
are larger than $O[(H_{\rm inf}/M)^2]$ because they arise with
coefficients of order $(v/M)^2$, where $v \gg H_{\rm inf}$ is the
scale in the scalar potential which gives rise to inflation:
$H_{\rm inf} \sim v^2/M_p$. However, most of these $(v/M)^2$
effects arise as modifications of the inflaton potential and so
represent a change in the connection between the slow-roll
parameters and the underlying string parameters, rather than an
observable deviation from the physical predictions of slow-roll
inflation themselves.
\end{enumerate}

It is clear that it is still early days for the exploration of the
implications of string theory for cosmology in the very early and
more recent universe. But the preliminary results are encouraging,
and the prospects remain bright for learning something interesting
about the physics of very high energies.

And perhaps the string-cosmology connection will prove to be even
more interesting once it is understood --- such as perhaps
providing an alternative to inflation \cite{Ekpyrosis}, by
providing accelerating universes through more exotic kinds of
stringy sources \cite{SBraneCosmo}, or in some other way we do not
yet anticipate. Let us hope so!

\section*{Acknowledgements}
I wish to thank the organizers of Moriond 2006 for arranging such
a pleasant and stimulating environment for discussions --- and
such excellent snow conditions --- as well as for their kind
invitation to speak. My understanding of the issues involved at
the interface of string theory and cosmology has benefited
immensely from interactions with many collaborators.
My research has been funded in part by McMaster University, the
Killam Foundation, and by the Natural Sciences and Engineering
Research Council of Canada.

\end{document}